\documentclass[11pt, twoside]{article}
\usepackage{amssymb, amsmath, amsfonts, latexsym, verbatim, mathrsfs}

\newtheorem{theorem}{Theorem}
\newtheorem{itlemma}{Lemma}[section]
\newtheorem{itproposition}[itlemma]{Proposition}
\newtheorem{itcorollary}[itlemma]{Corollary}
\newtheorem{itremark}[itlemma]{Remark}
\newtheorem{itremarks}[itlemma]{Remarks}
\newtheorem{itdefinition}[itlemma]{Definition}
\newtheorem{itexample}[itlemma]{Example}

\newenvironment{lemma}{\begin{itlemma}\rm}{\end{itlemma}} 
\newenvironment{remark}{\begin{itremark}\rm}{\end{itremark}} 
\newenvironment{remarks}{\begin{itremarks} \rm}{\end{itremarks}}
\newenvironment{corollary}{\begin{itcorollary}\rm}{\end{itcorollary}}
\newenvironment{proposition}{\begin{itproposition}\rm}{\end{itproposition}}
\newenvironment{definition}{\begin{itdefinition}\rm}{\end{itdefinition}}
\newenvironment{example}{\begin{itexample}\rm}{\end{itexample}}
\newenvironment{fact}{\noindent {\em Fact}. \ \ }{\hfill \medskip}
\newenvironment{claim}{\noindent {\em Claim}. \ \ }{\hfill \medskip}
\newcommand{\be}[1]{\begin{equation}\label{#1}}
\newcommand{\ee}{\end{equation}}
\newcommand{\bl}[1]{\begin{lemma}\label{#1}}
\newcommand{\br}[1]{\begin{remark}\label{#1}}
\newcommand{\brs}[1]{\begin{remarks}\label{#1}}
\newcommand{\bt}[1]{\begin{theorem}\label{#1}}
\newcommand{\bd}[1]{\begin{definition}\label{#1}}
\newcommand{\bp}[1]{\begin{proposition}\label{#1}}
\newcommand{\bc}[1]{\begin{corollary}\label{#1}}
\newcommand{\bfact}[1]{\begin{fact}\label{#1}}
\newcommand{\bex}[1]{\begin{example}\label{#1}}
\newcommand{\ec}{\end{corollary}}
\newcommand{\efact}{\end{fact}}
\newcommand{\eex}{\end{example}}
\newcommand{\el}{\end{lemma}}
\newcommand{\er}{\end{remark}}
\newcommand{\ers}{\end{remarks}}
\newcommand{\et}{\end{theorem}}
\newcommand{\ed}{\end{definition}}
\newcommand{\ep}{\end{proposition}}
\newcommand{\epr}{\end{proof}}
\newcommand{\bpr}{\begin{proof}}
\newcommand{\bcl}{\begin{claim}}
\newcommand{\ecl}{\end{claim}}

\newcommand{\bi}{\begin{itemize}}
\newcommand{\ei}{\end{itemize}}
\newcommand{\ben}{\begin{enumerate}}
\newcommand{\een}{\end{enumerate}}

\begin{document}

\title{Impact of positivity and complete positivity on
accessibility of Markovian dynamics}

\author{Raffaele Romano \footnote{Department of Mathematics,
Iowa State University, Ames, IA 50011, USA, rromano@iastate.edu.
Supported by NSF Career Grant, ECS0237925}}

\date{}

\maketitle

\begin{abstract}

\noindent We consider a two-dimensional quantum control system
evolving under an entropy-increasing irreversible dynamics in the
semigroup form. Considering a phenomenological approach to the
dynamics, we show that the accessibility property of the system
depends on whether its evolution is assumed to be positive or
completely positive. In particular, we characterize the family of
maps having different accessibility and show the impact of that
property on observable quantities by means of a simple physical
model.


\end{abstract}


\section{Introduction}

Irreversible dynamics appear in many fields of atomic and nuclear
physics~\cite{slich} and in quantum chemistry. They have great
relevance in quantum optics~\cite{loui,scull}, in statistical
mechanics~\cite{gardi} and in the description of continuously
measured systems~\cite{bragi}. They describe the time evolution of a
physical system in interaction with a second system (usually the
external environment). Under rather mild assumptions (as for example
a weak interaction between system and environment) these dynamics
are Markovian, that is they consist of semigroups of maps whose
generator for an $n$-level system has the standard form

\begin{equation}\label{koss}
    \dot{\rho} = -i [H, \rho] + \sum_{k,l = 1}^{n^2 - 1} c_{kl} \left(
    F_k \rho F_l^{\dagger} - \frac{1}{2} \left\{ F_l^{\dagger} F_k,
    \rho \right\} \right),
\end{equation}

\noindent where $\rho$ is the density matrix associated to the
system (i.e. a $n \times n$ positive matrix with unit trace), $H$ is
an Hermitian operator and the set $\{ F_k, k = 1, \ldots , n^2 -
1\}$ satisfies ${\rm Tr} F_k^{\dagger} F_l = \delta_{kl}$, $k,l = 0,
\ldots , n^2 - 1$ and $F_0 = I / \sqrt{n}$. The $(n^2 - 1) \times
(n^2 - 1)$ matrix $C$ with entries $c_{kl}$ is Hermitian; for
entropy increasing time evolutions it is real and
symmetric~\cite{alic,breu}.

From (\ref{koss}) we get $\rho_t = \gamma_t [\rho_{in}]$ where $\{
\gamma_t, t\geqslant 0 \}$ is a one-parameter semigroup of maps
describing the time evolution of the system, and $\rho_{in}$,
$\rho_t$ are the initial and final (at time $t$) states
respectively. These dynamics must fulfill some requirements
necessary for a consistent interpretation of the mathematical
formalism. In particular, at any time $t$ they must preserve the
positivity and the trace of the density matrix they act over. The
evolutions generated by (\ref{koss}) are automatically
trace-preserving, however some constraints on the coefficients
$c_{kl}$ have to be imposed in order to preserve the positivity of
$\rho_t$.

\begin{definition} \label{def1}
The map $\gamma_t$ is said to be {\it positivity-preserving} or
simply {\it positive} if and only if $\gamma_t [\rho] \geqslant 0$
$\forall \, \rho \geqslant 0$.
\end{definition}

\begin{definition} \label{def2}
The map $\gamma_t$ is said to be {\it completely positive} if and
only if $\gamma_t \otimes I_n$ is positive $\forall \, n \in \mathbb
N$, with $I_n$ the $n$-dimensional identity.
\end{definition}

By definition, complete positivity implies positivity. It is usually
believed (or assumed) that the evolutions generated by (\ref{koss})
should be completely positive rather than simply
positivity-preserving~\cite{alic,breu}. In fact, whereas
positivity-preserving maps guarantee that $\rho_t$ remains positive
at any time, the stronger property of complete positivity is
necessary to preserve the positivity of states initially entangled
with the environment (or part of it)~\cite{alic,breu,bena1,bena2}.

\begin{remark} \label{rem1}
The dynamics generated by (\ref{koss}) is completely positive if and
only if $C \geqslant 0$~\cite{alic,breu}.
\end{remark}

It is worth to consider that equation (\ref{koss}) is the Markovian
approximation of an exact, irreversible dynamics. This dynamics, for
an open system, is given by $\rho_t = {\rm Tr}_E ({\cal U}_t
\rho_0)$, where $\rho_0$ is the initial state of the composite
system (open system + environment), ${\cal U}_t$ its unitary time
evolution and ${\rm Tr}_E$ the partial trace over the environment
degrees of freedom. For an uncorrelated initial state, $\rho_0 =
\rho_{in} \otimes \rho_E$ (where $\rho_E$ is a reference state in
the environment) this dynamics is completely positive. Indeed it is
the composition of three completely positive maps, the expansion
$\rho_{in} \rightarrow \rho_{in} \otimes \rho_E$, the unitary
evolution ${\cal U}_t$ and the partial trace ${\rm Tr}_E$. From this
point of view, the preservation of complete positivity is a rather
natural assumption once a Markovian approximation of the dynamics is
assumed\footnote{This argument does not hold for a non-separable
initial state $\rho_0$ (for more details, see ref.\cite{pech},
\cite{alic3})}.

Although Markovian approximations leading to completely positive
time evolutions are always possible~\cite{davi} and, following the
previous discussion, desirable, they are not universally adopted:
sometimes the simple positivity is asked
for~\cite{sua,budi,laird,gasp1,gasp2}.

There are two approaches leading to the generator (\ref{koss}): an
axiomatic and a constructive one~\cite{alic2}. In the first case
both the form (\ref{koss}) and the complete positivity of the
resulting time evolution are assumed (this is the usual approach in
the quantum theories of information and computation). In the latter
case, (\ref{koss}) is derived as the Markovian approximation of a
generalized master equation, and complete positivity or positivity
are imposed as a second step. This procedure leads to
phenomenological coefficients $c_{kl}$ that are fixed but in general
unknown, since they embody the microscopic details of the
interaction between the open system and its external environment,
usually not accessible. In general, these coefficients are not all
independent and the assumptions of positivity and complete
positivity of the dynamics add further constraints.

Complete positivity is physically motivated introducing the
correlations of the considered system with an arbitrary external
system. Moreover it implies a hierarchy on the relaxing times of the
elements of $\rho_t$ (the diagonal with respect to the off-diagonal
ones) absent if the weaker property of positivity is asked
for~\cite{alic}. For these reasons in the previously cited works
simply positivity-preserving maps are preferred to completely
positive ones: to their authors, complete positivity appears as an
artificial mathematical request affecting too strongly the physical
behavior of the system.

In the following we will assume the constructive point of view for a
Markovian dynamics and address the problem of the choice between
positivity and complete positivity from a control theoretical point
of view. We will show that for some families of dynamics this choice
leads to different accessibility properties and, in turn, to
detectable differences in some observable quantity of the system.

We limit our attention to $2$-level systems evolving under entropy
increasing irreversible evolutions. This case describes many
interesting physical apparatuses, moreover it presents a simple
characterization of positive maps. We choose as basis operators the
Pauli matrices, $F_k = \sigma_k$, $k = 1, 2, 3$ and $\sigma_0 = I$,
and write a coherent vector representation of (\ref{koss}):

\begin{equation}\label{cohvec}
    \dot{\vec{\rho}} = {\cal L} \vec{\rho} = -(\cal{H} + \cal{D}) \vec{\rho},
\end{equation}

\noindent where we defined the vector $\vec{\rho} = (\rho_1, \rho_2,
\rho_3)$ with $\rho_k = {\rm Tr} (\rho \sigma_k) /2$, $k = 1, 2, 3$.
Since the time evolution is trace-preserving, $\rho_0 = {\rm Tr}
(\rho \sigma_0) /2 = 1 / 2$ is a constant. Positivity of $\rho$
translate to $\parallel \vec{\rho}
\parallel ^2 = \langle \vec{\rho}, \vec{\rho} \rangle \leqslant 1 / 4$,
defining the Bloch sphere ${\mathbb B}_0(1/2)$. The $3 \times 3$
real matrices ${\cal H}$ and ${\cal D}$ are skew-symmetric,
respectively symmetric and represent the Hamiltonian (or coherent)
contribution and the dissipative one.



\begin{remark} \label{rem2}
The entropy-increasing dynamics generated by (\ref{koss}) for a
two-level system is positivity-preserving if and only if ${\cal D}
\geqslant 0$~\cite{alic}.
\end{remark}

\noindent So, in this framework both positivity and complete
positivity of the dynamics are fully characterized by the positivity
of $3 \times 3$ matrices.


We further assume that the dynamics can be externally modified by
$m$ control functions $u_1, \ldots , u_m$, affecting the Hamiltonian
contribution, that is $H = H_0 + u_1 H_1 + \ldots + u_m H_m$. In the
coherence vector formalism,

\begin{equation}\label{control}
    \dot{\vec{\rho}} = -({\cal H}_0 + \sum_{i = 1}^m u_i {\cal H}_i + {\cal D})
    \vec{\rho}, \quad \vec{\rho}(0) = \vec{\rho}_i.
\end{equation}


\noindent Integrating (\ref{control}) we get a multi-parameter
semigroup of time evolutions. Our aim is to characterize the cases
where the controllability and accessibility properties can be
different under the requests of positivity or complete positivity.

We say that $\vec{\rho}^{\,\prime}$ can be reached from $\vec{\rho}$
at time $t$ if there exist some controls $u_i$, $i = 1, \ldots , m$
such that the time evolution generated by (\ref{control}) steers
$\vec{\rho}$ to $\vec{\rho}^{\,\prime}$ at time $t$. The set of all
$\vec{\rho}^{\,\prime}$ which are attainable from $\vec{\rho}$ at
time $t$ is denoted by ${\cal R}(\vec{\rho}, t)$. The following
definitions and properties are standard in Control
Theory~\cite{jurd}.

\begin{definition} \label{def3}
The {\it reachable set from $\vec{\rho}_i$ at time T} for the system
(\ref{control}) is given by
\begin{equation*}
{\cal R}(\vec{\rho}_i, T) = \bigcup_{0 \leqslant t \leqslant T}
{\cal R}(\vec{\rho}_i,t).
\end{equation*}
\end{definition}

\begin{definition} \label{def4}
The {\it reachable set from $\vec{\rho}_i$} for the system
(\ref{control}) is given by \begin{equation*} {\cal R}(\vec{\rho}_i)
= \bigcup_{t \geqslant 0} {\cal R}(\vec{\rho}_i,t).
\end{equation*}
\end{definition}

\noindent These sets depend on the particular choice of the initial
state $\vec{\rho}_i$.




\begin{definition} \label{def5}
System (\ref{control}) is {\it accessible} if and only if ${\cal
R}(\vec{\rho}_i, T)$ contains nonempty open sets of ${\mathbb B}_0
(1 / 2)$ $\forall \, T > 0$.
\end{definition}

From a physical point of view, accessibility means that the system
can be driven in every direction in the state space. If a system is
not accessible, there are forbidden directions in the state space
for the time evolution of the initial state. This affects the
reachable sets in the two cases. However, the evaluation of the
reachable sets is usually more difficult than the study of the
accessibility property. In fact, accessibility can be expressed by a
simple property, the so-called Lie algebra rank condition (LARC).

\begin{remark} \label{rem3}
System (\ref{control}) is accessible if and only if $Lie({\cal H}_0
+ {\cal D}, {\cal H}_1, \ldots, {\cal H}_m))$ is transitive on
${\mathbb B}_0 (1 / 2)$, that is $Lie({\cal H}_0 + {\cal D}, {\cal
H}_1, \ldots, {\cal H}_m) = \mathfrak{gl}(3, \mathbb{R})$ or
$\mathfrak{sl}(3, \mathbb{R})$.
\end{remark}

\begin{definition} \label{def6}
System (\ref{control}) is {\it controllable} if and only if $\forall
\, (\rho_i, \rho_f) \in {\mathbb B}_0(1 / 2) \times {\mathbb B}_0 (1
/ 2)$ there is a set of controls $u_1, \ldots , u_m$ such that
$\vec{\rho}(0) = \vec{\rho}_i$ and $\vec{\rho} (t) = \vec{\rho}_f$
for some $t\geqslant 0$.
\end{definition}

This means that we can steer every initial state to an arbitrary
final state using a suitable succession of controls. Otherwise said,
${\cal R}(\vec{\rho}_i) = {\mathbb B}_0 (1/2)$ for every initial
state $\vec{\rho}_i$: we can arbitrarily move into the state space
using the control functions.

The controllability properties of quantum mechanical systems have
been characterized for both closed systems (see~\cite{dale} and
references therein) and open ones. In particular, in the latter case
both accessibility and controllability properties have been fully
investigated in~\cite{alta} under the assumption of complete
positivity. In particular, the system is not controllable since $d /
dt \parallel \vec{\rho} \parallel ^2 = 2 (\dot{\vec{\rho}},
\vec{\rho}) = - ({\cal D}\vec{\rho},\vec{\rho}) \leqslant 0$. This
condition holds true even if we assume positivity instead of
complete positivity, therefore the results concerning
controllability are the same under the constraints of positivity and
complete positivity. However, the system could be accessible or not
in the two cases.



As a first step, in Section \ref{sec1} we discuss how the requests
of positivity and complete positivity affect the space of parameters
associated to a dissipative dynamics. In particular, Theorem
\ref{theo1} characterizes those dynamics for which the constraints
of positivity and complete positivity lead to different - but not
trivial - generators, a necessary condition for different
accessibility properties. In Section \ref{sec2} we consider this
class of maps and, using the LARC, we show that Theorem \ref{theo1}
is not a sufficient condition for a different accessibility. After
exploring a concrete example of control system belonging to this
family, we discuss our results.


\section{Necessary conditions for different accessibility properties}
\label{sec1}

Given a generator in the form (\ref{koss}), we assume that the
entries of $C$ are either independent or linearly dependent unknown
phenomenological parameters. Neglecting upper bounds on their
values\footnote{This assumption won't affect our results}, there is
a one-to-one correspondence between the matrices $C$ and the linear
spaces ${\cal V} \subseteq {\mathbb R}^6$. The number of independent
entries of $C$ is given by $n = dim \, {\cal V}$.



The requests of positivity and complete positivity define two convex
cones in the linear space of $3 \times 3$ real symmetric matrices,
${\mathscr C}_p$ (given by ${\cal D} \geqslant 0$) and ${\mathscr
C}_{cp}$ ($C \geqslant 0$). Since complete positivity is a stronger
property than positivity, ${\mathscr C}_{cp} \subset {\mathscr
C}_p$. We consider ${\mathscr S}_p = {\cal V} \cap {\mathscr C}_{p}$
and ${\mathscr S}_{cp} = {\cal V} \cap {\mathscr C}_{cp}$, the sets
of matrices $C$ restricted by the conditions of positivity or
complete positivity, and ${\mathscr S}_{cp} \subseteq {\mathscr
S}_p$. An arbitrary ${\cal V} \in {\mathbb R}^6$ will fit one of the
following cases ($0$ is the null matrix):

\begin{enumerate}
    \item ${\mathscr S}_{p} = \{0\}$ and ${\mathscr S}_{cp} =
    \{0\}$;
    \item ${\mathscr S}_{p} \ne \{0\}$ and ${\mathscr S}_{cp} =
    \{0\}$,
    \begin{enumerate}
        \item ${\mathscr S}_{p} \subset \partial
        {\mathscr C}_{p}$,
        \item ${\mathscr S}_{p} \nsubseteq \partial
        {\mathscr C}_{p}$;
    \end{enumerate}
    \item ${\mathscr S}_{p} \ne \{0\}$ and ${\mathscr S}_{cp}
    \ne \{0\}$,
    \begin{enumerate}
        \item ${\mathscr S}_{p} \subset \partial {\mathscr C}_{p}$
        and ${\mathscr S}_{cp} \subset \partial {\mathscr C}_{cp}$,
        \item ${\mathscr S}_{p} \nsubseteq \partial
        {\mathscr C}_{p}$ and ${\mathscr S}_{cp} \subset \partial
        {\mathscr C}_{cp}$,
        \item ${\mathscr S}_{p} \nsubseteq \partial
        {\mathscr C}_{p}$ and ${\mathscr S}_{cp} \nsubseteq
        \partial {\mathscr C}_{cp}$.
    \end{enumerate}
\end{enumerate}

In case $1$ any dynamics does not preserve the positivity of the
states over which it acts, in case $2$ there are not completely
positive dynamics, while in case $3$ both positive and completely
positive time evolutions can be obtained choosing the entries of
$C$.

The number of independent entries of $C$ after the requests of
positivity or complete positivity is equal to the dimension of the
smaller linear space containing the sets ${\mathscr S}_p$,
${\mathscr S}_{cp}$: ${\cal V}_{p} = span \{ {\mathscr S}_{p} \}$
and ${\cal V}_{cp} = span \{ {\mathscr S}_{cp} \}$. We define $n_p =
dim \, {\cal V}_p$ and $n_{cp} = dim \, {\cal V}_{cp}$. In general
${\cal V}_{cp} \subseteq {\cal V}_p$, then $n_{cp} \leqslant n_{p}
\leqslant n$. Different accessibility properties under the requests
of positivity and complete positivity are possible only if $n_{cp} <
n_p$. Otherwise, although the constraints between the dissipative
parameters are different in the two cases, we get the same Lie
algebra and then, via the LARC, accessibility is independent on the
constraint imposed.

In order to compute $n_p$ and $n_{cp}$ in every case, we derive some
relevant properties of the tangent space at the boundary of the cone
of real, positive and symmetric $3 \times 3$ matrices, $\partial
{\mathscr C}$. The inner product in ${\mathbb R}^6$ will be denoted
by $\langle \, \cdot \, , \, \cdot \, \rangle$.

\begin{lemma} \label{lem1}
The $5$-dimensional linear space ${\cal T}$ is the tangent space in
some $P \in \partial {\mathscr C}$ if and only if $\exists \,
\vec{v} \in \mathbb{R}^3$, $\vec{v} \ne \vec{0}$, such that $\langle
\vec{v}, T \vec{v} \rangle = 0$ $\forall \, T \in {\cal T}$.

\noindent {\it Proof:}\, The boundary $\partial {\mathscr C}$ is
given by the set of positive matrices with at least one vanishing
eigenvalue. In fact, since $P$ is real and symmetric, it can be put
in diagonal form by means of an orthogonal transformation, that is
$P ={\cal O} \tilde{P} {\cal O}^T$, ${\cal O} \in O(3)$ and
$\tilde{P}$ a diagonal matrix with $(\tilde{P})_{ii} = 0$ for some
$i \in \{1, 2, 3\}$. Except for the null matrix, $\partial {\mathscr
C}_p$ is a smooth $5$-dimensional manifold and its tangent spaces
are $5$-dimensional linear spaces. In particular, ${\cal T} = {\cal
O} \tilde{\cal T} {\cal O}^T$ where $\tilde{\cal T}$ is the tangent
space in $\tilde{P}$. Since $(\tilde{P})_{ii} = 0$ for some $i \in
\{1, 2, 3\}$, it follows $(\tilde{T})_{ii} = 0$ $\forall \,
\tilde{T} \in \tilde{\cal T}$. Consequently $\langle \vec{e}_i,
\tilde{T} \vec{e}_i \rangle = 0$ $\forall \, \tilde{T} \in
\tilde{\cal T}$, where $\vec{e}_i$ is defined by $(\vec{e}_i)_j =
\delta_{ij}$, and finally $\langle \vec{v}, T \vec{v} \rangle = 0$
$\forall \, T \in {\cal T}$, with $\vec{v} = {\cal O} \vec{e}_i$.
\hfill $\square$
\end{lemma}

\begin{lemma} \label{lem2}
If ${\cal T}$ is a tangent space of the manifold $\partial {\mathscr
C}$, then $span \{\partial {\mathscr C}_p \cap {\cal T}\}$ is a
$3$-dimensional linear subspace of ${\cal T}$.

\noindent {\it Proof:}\, The linear spaces $span \{\partial
{\mathscr C} \cap {\cal T}\}$ and $span \{\partial {\mathscr C} \cap
\tilde{\cal T}\}$ are isomorphic. Given $\tilde{T} \in
\partial {\mathscr C} \cap \tilde{\cal T}$, by Lemma \ref{lem1} and
since $\tilde{T} \geqslant 0$ the $i$-th row and column of
$\tilde{T}$ must vanish. Then $span \{\partial {\mathscr C} \cap
\tilde{\cal T}\}$ is $3$-dimensional and the thesis follows. \hfill
$\square$
\end{lemma}

\noindent Given an arbitrary $C$, the matrix ${\cal D}$ can be
expressed in terms of its entries:

\begin{equation}\label{CD}
   C = \left(
    \begin{array}{ccc}
    c_{11} & c_{12} & c_{13} \\
    c_{12} & c_{22} & c_{23} \\
    c_{13} & c_{23} & c_{33} \\
    \end{array}
    \right), \quad
    {\cal D} = 2 \left(
    \begin{array}{ccc}
    c_{22} + c_{33} & -c_{12} & -c_{13} \\
    -c_{12} & c_{11} + c_{33} & -c_{23} \\
    -c_{13} & -c_{23} & c_{11} + c_{22} \\
    \end{array}
    \right)
\end{equation}

\noindent that is ${\cal D} = 2 (I\,{\rm Tr} \, C - C)$. Since there
is an isomorphism between unitary transformations of the Pauli
matrices, $\tilde{\sigma}_i = {\cal U} \sigma_i {\cal U}^{\dagger}$
with ${\cal U} \in U(2)$, and orthogonal transformations of $C$ (and
${\cal D})$, $\tilde{C} = {\cal O}^T C {\cal O}$ (and $\tilde{\cal
D} = {\cal O}^T {\cal D} {\cal O}$) with ${\cal O} \in O(3)$, the
orthogonal transformations in Lemma \ref{lem1} are related to
unitary changes of basis. In particular, the relation between the
entries of $C$ and ${\cal D}$ is basis independent since
$\tilde{\cal D} = 2 (I\,{\rm Tr} \, \tilde{C} - \tilde{C})$. Notice
also that whenever ${\cal V}$ is a subset of the boundary of some
cone, it is a linear subspace of a tangent space to this boundary.
Then, following Lemmas \ref{lem1} end \ref{lem2} and considering the
expressions (\ref{CD}), $n_{p}$ and $n_{cp}$ can be evaluated in
every case. They are listed in the following table:

\begin{center}
\begin{tabular}{c|c|c|c} \label{tab1}
  case & $n$ & $n_{p}$ & $n_{cp}$ \\ \hline
  1 & $\leqslant 5$ & 0 & 0 \\
  2(a) & $\leqslant 5$ & $\leqslant 3$ & 0 \\
  2(b) & $\leqslant 5$ & $n$ & 0  \\
  3(a) & $\leqslant 4$ & $\leqslant 2$ & 1 \\
  3(b) & $\leqslant 5$ & $n$ & $\leqslant 3$ \\
  3(c) & $\leqslant 6$ & $n$ & $n$ \\
\end{tabular}
\end{center}

The requests of positivity and complete positivity produce linear
spaces ${\cal V}_{p}$ and ${\cal V}_{cp}$ lower dimensional than
${\cal V}$ in two cases. Either ${\cal V}$ is a linear subspace of a
tangent space to the boundaries of the respective cones, or it
intersects these cones only in the null matrix $0$. The only non
trivial cases admitting both positive and completely positive
dissipative contributions with $n_p > n_{cp}$ are 3(a) and 3(b). For
later reference, we present here a necessary and sufficient
condition for the form of the generators of these dynamics.

\begin{theorem} \label{theo1}
Given a generic $C$ in (\ref{koss}), define $K = span \{\vec{w} \in
{\mathbb R}^3 \vert \langle \vec{w}, C \vec{w} \rangle = 0 \}$. Then
$n_p > n_{cp}\ne 0$ if and only if one of the two following
conditions is satisfied:
\begin{enumerate}
    \item $dim \, K = 1$ and $\forall \, \vec{w} \in K$, $\vec{w} \ne 0$ we have $C \vec{w} \ne
    0$;
    \item $dim \, K = 2$ and $\exists \, \vec{w}_1, \vec{w}_2 \in K$
    such that $\langle \vec{w}_1, \vec{w}_2 \rangle = 0$ and $\langle \vec{w}_1, C \vec{w}_2 \rangle \ne 0$.
\end{enumerate}
\end{theorem}

\noindent {\it Proof:}\, As previously stated, $n_p > n_{cp}\ne 0$
if and only if we are in case 3(a) or 3(b).

\noindent In case 3(a), by Lemma \ref{lem1}, $\exists \, \vec{v} \in
\mathbb{R}^3$, $\vec{v} \ne \vec{0}$, such that $\langle \vec{v},
{\cal D} \vec{v} \rangle = 0$ or, in a suitable basis, $(\tilde{\cal
D})_{ii} = 0$ for some $i \in \{1, 2, 3\}$ (whit $\vec{v} = {\cal O}
\vec{e}_i$ and $\tilde{\cal D} = {\cal O}^T {\cal D} {\cal O}$) and
thus $\tilde{c}_{jj} = 0$ for $j \ne i$. Therefore $dim \, K = 2$.
Moreover $n_p \ne n_{cp}$ if and only if $\tilde{c}_{jk} \ne 0$,
with $j \ne k$ and $j, k \ne i$. Defining $\vec{w}_{1,2} = {\cal O}
\vec{e}_{j,k}$ we have $\langle \vec{w}_1, \vec{w}_2 \rangle = 0$
and $\langle \vec{w}_1, C \vec{w}_2 \rangle \ne 0$.

\noindent In case 3(b), $\exists \, \vec{v} \in \mathbb{R}^3$,
$\vec{v} \ne \vec{0}$, such that $\langle \vec{v}, C \vec{v} \rangle
= 0$ or, arguing as before, $\tilde{c}_{ii} = 0$ for some $i \in
\{1, 2, 3\}$. Then $dim \, K \geqslant 1$. Necessary and sufficient
condition for $n_p \ne n_{cp}$ is $\tilde{c}_{ij} \ne 0$ for some $j
\ne i$. In particular, if $dim \, K = 1$ define $\vec{w} = {\cal O}
\vec{e}_{j}$; if $dim \, K = 2$ define $\vec{w}_{1,2} = {\cal O}
\vec{e}_{j,k}$ with $j \ne k$ and $j, k \ne i$. Since complete
positivity implies $\tilde{c}_{ij} = 0$ for $j \ne i$, the thesis
follows. \hfill $\square$


\section{Accessibility for positive and completely positive maps}
\label{sec2}

In this section we study the accessibility properties of some
selected evolutions fitting the conditions expressed by Theorem
\ref{theo1}. We limit our attention to a single control $u$
switching on/off the Hamiltonian part,

\begin{equation}\label{piec}
    {\cal L}(u) =
    -(u {\cal H} + {\cal D}),
\end{equation}

\noindent with $u = 0, 1$. The more general dissipative contribution
in (\ref{cohvec}), restricted by the requests of positivity or
complete positivity, up to an orthogonal transformation is given by

\begin{equation}\label{dpdcp}
    {\cal D}_p = 2 \begin{pmatrix}
      c_{22} & -c_{12} & -c_{13} \\
      -c_{12} & c_{11} & -c_{23} \\
      -c_{13} & -c_{23} & c_{11} + c_{22} \\
    \end{pmatrix},\,\,\,
{\cal D}_{cp} = 2 \begin{pmatrix}
      c_{22} & -c_{12} & 0 \\
      -c_{12} & c_{11} & 0 \\
      0 & 0 & c_{11} + c_{22} \\
    \end{pmatrix}
\end{equation}

\noindent with $(c_{13}, c_{23}) \ne (0, 0)$. To simplify the
computations, we assume $c_{12} = c_{13} = 0$. The Hamiltonian
contribution is given by the skew-symmetric matrix

\begin{equation}\label{h}
    {\cal H} = 2 \begin{pmatrix}
      0 & h_3 & -h_2 \\
      -h_3 & 0 & h_1 \\
      h_2 & -h_1 & 0 \\
    \end{pmatrix},
\end{equation}

\noindent where $h_k = {\rm Tr} (H \sigma_k)/2$, $k = 1 ,2 ,3$, and
$H$ in (\ref{koss}). Without loss of generality, we choose either
$h_1 \ne 0$ and $h_2 = h_3 = 0$, or $h_1 = h_2 = 0$ and $h_3 \ne 0$.

\noindent The Lie algebras generated in the two cases are denoted by
${\cal A}_p$ and ${\cal A}_{cp}$:

\begin{eqnarray}
  {\cal A}_p &=& Lie ({\cal D}_p, {\cal H}+{\cal D}_p) \\
  \nonumber{\cal A}_{cp} &=& Lie ({\cal D}_{cp}, {\cal H}+{\cal D}_{cp}).
\end{eqnarray}

A convenient orthogonal basis for the algebra of $3 \times 3$ real
matrices, $\mathfrak{gl}(3, {\mathbb R})$, is given by the matrices $E_{ij}$,
$i, j = 1, 2, 3$ defined by $(E_{ij})_{kl} = \delta_{ik}
\delta_{jl}$. In this basis we have

\begin{enumerate}
    \item for $h_1 = h_2 = 0$, $h_3 \ne 0$,
    \begin{description}
        \item[] for $c_{11} = c_{22}$, ${\cal A}_{p} = \mathfrak{gl}(3, {\mathbb
        R})$ and ${\cal A}_{cp} = span \{ E_{21} - E_{12}, E_{11} + E_{22} + 2
        E_{33}\}$,
        \item[] for $c_{11} \ne c_{22}$, ${\cal A}_{p} = \mathfrak{gl}(3, {\mathbb
        R})$ and ${\cal A}_{cp} = span \{ E_{12}, E_{21}, E_{11} - E_{22}, E_{22} + E_{33}\}$;
    \end{description}
    \item for $h_2 = h_3 = 0$, $h_1 \ne 0$,
    \begin{description}
        \item[] ${\cal A}_{p} = {\cal A}_{cp} = span \{ E_{23}, E_{32}, E_{22} + E_{33},
        2 c_{11} E_{22} + c_{22} (E_{11} + E_{22}) \}$.
    \end{description}

\end{enumerate}

Therefore the system is accessible for positive maps but not for
completely positive ones in case 1 and it is never accessible in
case 2. Thus different accessibility properties under the requests
of positivity and complete positivity are possible, moreover Theorem
\ref{theo1} is not a sufficient condition for them.


\section{Accessibility for a spin in a stochastic magnetic field}
\label{sec3}

Consider a spin evolving under the action of a stochastic magnetic
field $\vec{B}(t) = \langle \vec{B}(t) \rangle + \vec{\beta} (t)$,
where $\langle \vec{B} (t) \rangle = (0, 0, B_3)$ is the
time-independent average and $\vec{\beta} (t) = (\beta_1 (t), 0,
\beta_3 (t))$ is a two-component stochastic part, with $\langle
\vec{\beta} (t) \rangle = (0, 0, 0)$. This configuration can be
obtained via a perfect shielding of the $y$ component of the
magnetic field. The control consists in switching on/off $B_3$ while
the stochastic part is not affected by it. The two-time correlations
of the stochastic components are given by

\begin{equation}\label{corr}
    W_{ij} (t) = \langle \beta_i (t) \beta_j (0) \rangle,
\end{equation}

\noindent entries of the real, positive definite covariance matrix
$W(t)$. We assume that both diagonal and off-diagonal correlations
functions are non-vanishing.



We can describe the time evolution of the spin system introducing a
stochastic $2 \times 2$ density matrix $\rho_s$ satisfying the
semi-classical Liouville-Von Neumann equation

\begin{equation}\label{sc}
    \dot{\rho}_s (t) = -i [B_3 \sigma_3 + \vec{\beta} (t) \cdot \vec{\sigma}, \rho_s (t)]
\end{equation}

\noindent where $\vec{\sigma} = (\sigma_x, \sigma_y, \sigma_z)$.
Using the so-called convolutionless approach~\cite{budi} it is
possible to obtain the time evolution for the density matrix
averaged over the stochastic component, $\rho (t) = \langle \rho_s
(t) \rangle$:

\begin{equation}\label{ca}
    \dot{\rho} (t) = -i [B_3 \sigma _3, \rho (t)] - \sum_{k,l = 1}^3
    \hat{c}_{kl} (t) [\sigma_k,[\sigma_l, \rho(t)]]
\end{equation}

\noindent where

\begin{equation}\label{ckl}
    \hat{c}_{kl} (t) = \sum_{j = 1}^3 \int_0^t W_{kj} (s) U_{jl} (-s) ds
\end{equation}

\noindent and the $U_{jl}$ are the matrix elements of

\begin{equation}\label{u}
    U (t) = \left(%
\begin{array}{ccc}
  \cos{2B_3 t} & - \sin{2B_3 t} & 0 \\
  \sin{2B_3 t} & \cos{2B_3 t} & 0 \\
  0 & 0 & 1 \\
\end{array}%
\right)
\end{equation}

A Markovian approximation is justified whenever the coupling between
the spin system and the external stochastic field is weak. It
corresponds to neglecting the memory effects in (\ref{sc}), in
practice $t \rightarrow + \infty$ in (\ref{ckl}) and we obtain a
time-independent Lindblad generator. The coefficients $c_{kl} =
\hat{c}_{kl} + \hat{c}_{lk}$ define the matrix

\begin{equation}\label{newc}
    C = \left(%
\begin{array}{ccc}
  c_{11} & c_{12} & c_{13} \\
  c_{12} & 0 & c_{23} \\
  c_{13} & c_{23} & c_{33} \\
\end{array}%
\right)
\end{equation}

\noindent where

\begin{eqnarray}
\nonumber  c_{11} &=& 2 \int_0^{+\infty} W_{11} (s) \cos{(2B_3
  s)} ds \\
\nonumber  c_{12} &=& \int_0^{+\infty} W_{11} (s) \sin{(2B_3
  s)} ds \\
  c_{13} &=& \int_0^{+\infty} W_{13} (s) (\cos{(2B_3
  s) + 1}) ds \\
\nonumber  c_{23} &=& \int_0^{+\infty} W_{13} (s) \sin{(2B_3
  s)} ds \\
\nonumber  c_{33} &=& 2 \int_0^{+\infty} W_{33} (s) ds.
\end{eqnarray}

\noindent In the coherence vector representation (\ref{ca}) becomes
$\dot{\vec{\rho}} = - ({\cal H} + {\cal D}) \vec{\rho}$ with


\begin{equation}\label{matr}
    {\cal H} = 2 \left(%
\begin{array}{ccc}
  0 & B_3 + \omega_{3} & \omega_{2} \\
  -B_3 - \omega_{3} & 0 & \omega_{1} \\
  - \omega_{2} & - \omega_{1} & 0 \\
\end{array}%
\right), \quad \quad
    {\cal D} = 2 \left(%
\begin{array}{ccc}
  c_{33} & - c_{12} & - c_{13} \\
  - c_{12} & c_{11} + c_{33} & - c_{23} \\
  - c_{13} & - c_{23} & c_{11} \\
\end{array}%
\right),
\end{equation}

\noindent where

\begin{eqnarray}
\nonumber  \omega_{1} &=& c_{23} \\
  \omega_{2} &=& \int_0^{+\infty} W_{13} (s) (\cos{(2B_3
  s) - 1}) ds \\
\nonumber  \omega_{3} &=& -c_{12} \\
\end{eqnarray}
The conditions of Theorem \ref{theo1} are fulfilled since $K = span
\{\vec{w} = (0, 1, 0)\}$, $dim \, K = 1$ and $C \vec{w} \ne 0$. The
constraint of positivity does not affect the dimension of the space
of parameters associated to this system whereas complete positivity
asks for $c_{12} = c_{23} = 0$. In the latter case, the stochastic
model is consistent only if the correlation functions are either
vanishing, or white noise. All the possible cases are listed in the
following table, together with the corresponding Lie algebras.

\begin{center}
\begin{tabular}{c|c|c|c} \label{tab2}
  $W_{11}$(t) & $W_{13}(t)$ & conditions & ${\cal A}_{cp}$ \\ \hline
  $0$ & $0$ & \begin{tabular}{c} $c_{11} = c_{12} = c_{13} = c_{23} = 0$ \\ $\omega_1 = \omega_2 = \omega_3 =
  0$ \end{tabular}
  & $span \{ E_{11} + E_{22},
  E_{12} - E_{21} \}$ \\ \hline
  $W_{11} \delta(t)$ & $0$ & \begin{tabular}{c} $c_{12} = c_{13} = c_{23} = 0$ \\ $\omega_1 = \omega_2 = \omega_3 =
  0$ \end{tabular} & $span \{ E_{11}, E_{12}, E_{21}, E_{22} \} \oplus span \{ E_{33} \}$ \\ \hline
  $W_{11} \delta(t)$ & $W_{13} \delta(t)$ & \begin{tabular}{c} $c_{12}=c_{23}=0$ \\ $\omega_1 = \omega_3 =
  0$ \end{tabular} & $\mathfrak{gl}(3, \mathbb{R})$ \\
\end{tabular}
\end{center}

If positivity is imposed, no assumptions have to be done on the
two-time correlation functions of the stochastic magnetic field and
the Lie algebra is ${\cal A}_p = \mathfrak{gl}(3, \mathbb{R})$.
Hence, the choice between positivity and complete positivity may
affect the accessibility property of the system, depending on the
assumptions on the correlations functions. The system is always
accessible under the request of positivity, whereas it is accessible
for a completely positive dynamics only if the stochastic magnetic
field is assumed to have white noise correlations.


\section*{Conclusions}

We discussed the impact of positivity and complete positivity of the
dynamics on controllability and accessibility of a two-dimensional
open system, evolving under a Markovian, entropy increasing time
evolution. Whereas controllability is insensitive of what property
is enforced, accessibility does. We gave in Theorem \ref{theo1} a
necessary condition for different accessibility and we discussed a
concrete example in Section \ref{sec3}.

We stress that we considered a phenomenological approach to the
dynamics of the system: the details of the interaction are unknown
and a model of time evolution is assumed. Positivity or complete
positivity are imposed after the Markovian approximation, leading to
some relations between the dissipative parameters that describe the
dynamics and eventually constraining the microscopic properties of
the surrounding. For example, in the case considered in Section
\ref{sec3} we observed that complete positivity is not compatible
with generic two-time correlation functions of the bath.

In general, there can be transitions allowed if positivity is
assumed and forbidden under complete positivity, since complete
positivity implies a hierarchy in the relaxation times of diagonal
and off-diagonal elements of the density matrix describing the
system, that in turn affects the reachable sets of the system.
However a different accessibility is a strongest condition, since it
implies that the dimensions (as manifolds) of the reachable sets are
different in the two cases.

Different accessibility properties have observable consequences. The
measurement outcomes of some selected physical quantities can
exhibit a dependence on whether positivity or complete positivity is
asked for. A simple example is the following. Consider the spin in
the stochastic magnetic field discussed in Section~\ref{sec3}.
Assume the initial state is polarized along the positive $x$
direction, that is $\rho (0) = \vert \uparrow_x \rangle \langle
\uparrow_x \vert$ or $\vec{\rho} (0) = (1/2, 0, 0)$. This state does
not exhibit any polarization in spin along the $z$ direction, $S_z$.
By means of the switching on/off control $u$ we want to get a (even
slightly) polarized state along the positive $z$ direction, that is
a state for which the average of $S_z$ does not vanish,

\begin{equation}\label{prob}
    \langle S_z (t)\rangle = {\rm Tr} \Bigl( \frac{1}{2} \sigma_z \rho(t)
    \Bigr) = \rho_3 (t) \ne 0.
\end{equation}

\noindent It is convenient to evaluate

\begin{equation}\label{der}
    \frac{d}{dt} \langle {S}_z (t)\rangle = \dot{\rho}_3 (t) =
    ({\cal L} \vec{\rho}(t))_3
\end{equation}

\noindent where ${\cal L} = - ({\cal H} + {\cal D})$. An explicit
computation using the results of the previous section shows that,
unless white noise correlations are assumed, for completely positive
maps $\frac{d}{dt} \langle {S}_z (0)\rangle = 0$ whereas for
positive maps $\frac{d}{dt} \langle {S}_z (0)\rangle \geqslant 0$.
Moreover, since

\begin{equation}\label{derhig}
    \frac{d^n}{dt^n} \langle {S}_z (t)\rangle = ({\cal L}^n
    \rho(t))_3
\end{equation}

\noindent and considering that $\langle S_z(0)\rangle = 0$, it
follows $\langle S_z(t)\rangle = 0$ for all time $t$ for completely
positive maps, whereas $\langle S_z(t)\rangle > 0$ at some $t$ for
positive maps. This behavior reflects the different accessibility
properties in the two cases.

A final remark concerns possible generalizations of this work. We
considered a Markovian approximation to the open system dynamics,
and we assumed an entropy-increasing time evolution. If we try to
relax these hypotheses, the characterization of controllability and
accessibility is very difficult. For Markovian, entropy-increasing
time evolutions these properties have been characterized
in~\cite{alta} under the request of complete positivity. It is not
trivial to generalize this result to simple positive maps since for
this class of maps there is not a satisfactory necessary and
sufficient condition expressing positivity, equivalent to
Remark~\ref{rem2}. More generally, it would be very difficult to
discuss the controllability and accessibility properties of the
exact (non Markovian) open system dynamics, since it is described by
a generalized master equation expressed by an integro-differential
equation.


\end{document}